# Statistical model for intermittent plasma edge turbulence

F. Sattin and N. Vianello

Consorzio RFX, Corso Stati Uniti 4,

35127 Padova, ITALY

**Abstract**

*The Probability Distribution Function of plasma density fluctuations at the edge of fusion devices is known to be skewed and strongly non-Gaussian. The causes of this peculiar behaviour are, up to now, largely unexplored. On the other hand, understanding the origin and the properties of edge turbulence is a key issue in magnetic fusion research. In this work we show that a stochastic fragmentation model, already successfully applied to fluid turbulence, is able to predict an asymmetric distribution that closely matches experimental data. The asymmetry is found to be a direct consequence of intermittency. A discussion of our results in terms of recently suggested BHP universal curve [S.T. Bramwell, P.C.W. Holdsworth, J.-F. Pinton, Nature (London)* **396**, *552 (1998)], that should hold for strongly correlated and critical systems, is also proposed*

# 1. Introduction

The investigation of the mechanisms underlying turbulence is a key topic in fusion research and, more broadly, plasma physics. In particular, thanks to the steady increase in computing power, the direct numerical solving of fluid or kinetic equations is more and more widespread. However, direct numerical simulations still have some drawbacks: first of all, one can hardly hope of tackling the full set of equations; rather, must truncate them by choosing *a priori* the relevant instabilities and this, of course, introduces some arbitrariness as well as loss of accuracy and predictive power. Second, turbulence is a mechanism involving widely differing spatial and temporal scales, and this is demanding for numerical computations. Finally, a difficult task with numerical simulations is how to abstract basic plasma properties from huge amount of data; that is, it is often difficult to grasp any intuitive picture of the problem at hand. For these reasons, phenomenological models are valuable: they guess from the outset some basic properties of the plasma, and complement them with an intuitive (but hopefully accurate) picture of the microscopic dynamics. The result is a model with good interpretive and predictive capabilities realized with an economy of concepts, mathematical and numerical machinery. The drawback being, usually, an agreement with experiment not exceedingly accurate.

Within this class of models, the best known is the approach based upon the Self Organized Criticality (SOC) paradigm, put forth by Carreras and co-workers (see, e.g., [1,2] for its description and application to plasmas), which has enjoyed widespread consideration. However, alongside with works supporting this theory, there is also evidence suggesting that SOC alone is too simplified a picture to account for all of the complexity shown in real plasmas. In particular, a key element required by SOC is self-similarity of plasma behaviour over the scale of lengths relevant for transport. Although this requisite was claimed to be satisfied in several devices [3], some other experimental results [4], supplemented also by simulations [5], suggest that this might not be the case, at least in a number of other cases. For this reason, alternate approaches have been suggested, e.g., based upon shell models [6].

A common finding of the studies devoted to the statistical properties of plasma turbulence is that many of its features are universal: independent of the device and, hence, of the details of the free energy driving the turbulence itself. This means that, in principle, one could develop a model based on very general principles, without any reference to specific mechanisms triggering and sustaining turbulence that, instead, are necessary in traditional fully numerical methods, and that can be specific to each setting. In this work we present an attempt of edge plasma turbulence modelling based on a statistical approach à la Kolmogorov. Intermittency-i.e. departure from self similarity-is naturally embodied into the model and, indeed, constitutes a fundamental part of it. The model is patterned after the paper by Portelli *et al* [7] (hereafter referred to as I), with small differences due to the different settings and quantities under observation. We will show, using this model, that the experimental Probability Distribution Functions (PDFs) for particle density fluctuations at the edge of a fusion device may be fairly well recovered.

## 2. Model for intermittent turbulence

The picture we are going to propose is the following: we suppose that particle density plays the role of a passive scalar advected by eddies of various sizes. Mixing processes make density almost uniform within each eddy, while two eddies may have even widely differing densities. A fragmentation process, which preserve the total particle density, does exists splitting larges eddies into smaller ones. Also, molecular processes contribute to further fragment eddies into a gas of independent particles and, in particular, such mechanisms are dominant at small lengths, i.e., no eddies are thought to exist below a given size (dissipative length) $\eta$. At the other extreme, there is instead the macroscopical scale *L* that sets the typical size over which actual measurements are performed. A comment is in order at this point. In fluid turbulence studies one has a clear-cut distinction between the medium (the fluid) and the passive scalar advected (usually, the flow velocity or acceleration). In this case, instead, the medium and the passive scalar do appear to coincide (eddies are made of particles).

Any experimental measurement of density results partially from this gas of independent eddies over all allowed sizes, and partially from the background of single-particle contributions. Experimentally, it has been found in a number of devices that the relative weights of the two contributions are comparable, when not in favour of the eddies (this statement addresses, to be precise, to particle flux, rather than particle density, but we are allowed to think that the results should be similar. See papers [3,4,8]). The independent-particle part alone would provide a purely Gaussian PDF of density fluctuations. Adding a substantial contribution from the coherent eddies will drastically modify the tails of this distribution. The central region around the maximum will be modified to a lesser extent, remaining approximately Gaussian. The degree of the perturbation and the width of this region depend on the relative weights of the two contributions as well as on the details on the statistics.

We use, as customary, a discrete set of scales $l$, labelled by the index $n$, $0 \leq n \leq \Lambda$; adopt the standard convention of a constant ratio between two adjacent scale lengths, $(l_{n-1}/l_n)^3 \equiv \rho$, and choose the largest and smallest scale equal to the macroscopical and dissipation length respectively: $l_0 \equiv L$, $l_\Lambda = \eta$, thus $\rho^\Lambda = (L/\eta)^3 \equiv Re$. With the latter definition we imply that $\rho^\Lambda$ plays the role of an effective Reynolds number.

Another key quantity is the instantaneous local density flux at scale $l_n$, $\tilde{\pi}_n(r,t)$ [units (length)$^{-3}\times$(time)$^{-1}$], i.e. the flux of matter that from largest scales flows into structures at scale $l_n$. Experimentally, only the largest scales are likely to be directly measurable. The smaller ones will be averaged by the measuring device and procedure. Hence, we cannot work directly with $\tilde{\pi}_n(r,t)$, rather the density flux must be averaged over the macroscopical observation volume:

$$\pi_n(t) = \frac{1}{L^3}\int d^3r\, \tilde{\pi}_n(r,t) \ . \qquad (1)$$

Notice that the mathematical averaging (1) has an exact experimental counterpart when the measurement is performed over a large volume, if compared with average eddies' size. This could be the case of density fluctuations measured through Neutral Beam Emission Spectroscopy. At the edge, measurements are performed through Langmuir probes, of moderate size even for plasma fluctuations. On the other hand, the typical measurement time cannot be made small with respect to all time scales. Hence, the average (1) is experimentally a time average, translated to a spatial one through ergodicity or Taylor's frozen turbulence hypothesis.

At each scale $l_n$, molecular processes will remove part of the particles from the coherent behaviour within the eddy. We designate by $\mu_0$ this rate of consumption per unit volume. This quantity, by definition, cannot depend from scale $l$.

We introduce now the excess instantaneous density: $\Delta N(t) = (N(t) - \overline{N})$. It is the difference between the instantaneous density at time $t$, $N(t)$, and its long-time averaged value $\overline{N}$. The excess density is determined by the net difference between source and loss terms at each length scale, summed over all scales:

$$\Delta N(t) = \tau \sum_n (\pi_n - \mu_0) \tag{2}$$

The time scale $\tau$, as remarked by Portelli *et al*, should be a purely macroscopical parameter, determined by density injection at the largest scales (fuelling mechanism). As such, it does not depend from scale *l*, nor it is likely it depends strongly from any microscopical detail within the present model.

Rigorously speaking, a fully consistent theory must provide the dynamics of $\pi_n(t)$ and, thus, of $\Delta N(t)$. Lacking of such a theory, we must disregard dynamics and turn to a statistical point of view, making the assumptions that the $(\pi_n - \mu_0)$ are statistically independent stochastic variables. Indeed, both $\pi_n$ and $\mu_0$ individually might be stochastic variables. However, even though $\mu_0$ may fluctuate, it depends from molecular processes, not from fluid ones; hence, we expect that rms($\delta\mu_0$) << rms($\delta\pi_n$), and $\mu_0$ will be considered as a constant offset.

It is Eq. (2) where intermittency comes into the model: in the standard Kolmogorov's K41 turbulence theory $\Delta N$ is not a stochastic variable but a constant. Since, by definition, its average value must be zero under stationary conditions, it must be null at all times as well. Hence the rhs of (2) must be postulated to be identically zero, too, while in our approach it is only in a statistical sense: $\langle\Delta N\rangle = 0$, but $\Delta N \neq 0$ almost always.

The PDF for *N*, $P(\Delta N)$, can be written starting from the product of PDFs for $(\pi_n - \mu_0)$:

$$\prod_n p(\pi_n - \mu_0) d(\pi_n - \mu_0) = p\left(\sum_n (\pi_n - \mu_0)\right) \prod_n d(\pi_n - \mu_0)$$

At this stage, we make the replacement of variables $\Delta N = \sum_n (\pi_n - \mu_0), x_1 = \pi_1 - \mu_0, ..., x_\Lambda = \pi_\Lambda - \mu_0$, thus apart for a trivial volume element, we may identify

$$P(\Delta N) = \prod_n p(\pi_n - \mu_0) \tag{3}$$

This is the standard problem of computing the PDF for a quantity sum of a finite number of other stochastic variables. The general solution is reviewed in a recent paper [9], although is fairly straightforward. We introduce the characteristic function

$\Psi(k)$: $\Psi(k) = \prod_n \tilde{p}_n(k)$ is the product of the Fourier transforms of $p(\pi_n - \mu_0)$. This yields

$$P(\Delta N) = \int_{-\infty}^{+\infty} \frac{dk}{2\pi} e^{ikN} \Psi(k) \qquad (4)$$

In order to simplify notation, here and henceforth we will normalize data to unity standard deviation: $\sigma(\Delta N) \equiv 1$.

The fundamental role is played by the PDF for the flux, $p(\pi_n)$. A lot of effort was devoted in turbulence studies to provide an analytical expression for this quantity, starting from the log-normal expression by Kolmogorov and Obukhov [10], to the log-Poisson by She and Levèque [11], just to mention some. Here, we note that Eq. (1) can be discretized into a sum over small equal-volume cells:

$$\pi_n(t) \approx \frac{\Delta V}{L^3} \sum_i \tilde{\pi}_n(\bar{r}_i, t) \qquad (5)$$

where $i$ is an index labelling one generic cell, $\bar{r}_i$ a point representative of the position of the cell, and $\Delta V$ the volume of the cell. Each cell may be given the size of the eddy at that scale: $\Delta V = l^3$. Hence, the total number of cells is given at each scale by $\nu(l_n) \approx [(L/l_n)^3] = [\rho^n]$ (the square brackets [...] stand for the integer part). We carry further the statistical view, and consider the $\tilde{\pi}_n(\bar{r}_i, t)$'s as $\nu$ stochastic variables. They are, by definition, positive-definite quantities: $\tilde{\pi}_n(\bar{r}_i, t) \equiv Z_i^2$. We can make just a few statements about the stochastic variables $Z_i$'s: I) they are indentical statistically independent variables; II) the average value of $\tilde{\pi}_n$ (hence $Z^2$) must coincide with $\mu_0$. We expect PDF($\tilde{\pi}_n$) also to be a reasonably well behaved function, vanishing to infinity and at $\tilde{\pi}_n = 0$ (by continuity, taking into account that negative values are not permitted). Finally, dealing with macroscopic systems, it is reasonable to assume PDF($\tilde{\pi}_n$) to have a single maximum, practically identical with the average value $\mu_0$. All these requisites are fulfilled by a chi-squared PDF for $Z_i^2$ or, in other terms, by a normal distribution for $Z_i$:: $P(Z_i) \propto \exp(-Z_i^2/(2\sigma_Z^2))$. Hence, we may write, using (5) together with the relative independence of variables $\tilde{\pi}$:

$$p(\pi_n) = p\left(\sum_i \tilde{\pi}_n(i)\right) = \prod_i p(\tilde{\pi}_n(i)) = \left(\frac{\nu}{\mu_0}\right)^\nu \frac{\pi_n^{\nu-1}}{\Gamma(\nu)} \exp\left(-\nu \frac{\pi_n}{\mu_0}\right) \qquad (6)$$

where we have already taken into account that the average value <$\pi_n$> = $\mu_0$ . The result (6) is a textbook exercise of composition of probability densities; the r.h.s. is a Gamma (or $\chi^2$) PDF.

This is the same expression used in I, guessed there just on the basis of the nice fit with experimental data. Here, we are providing also some theoretical ground for it: even though it must remain clear that ours is not a first-principle derivation, and there is some amount of arbitrariness, we think we have provided some sound reasons to suspect that Eq. (6) is a valid candidate to the true PDF.

Since the PDF of $\pi_n(t)$ plays a major role, we spend some more words about it. We found that it is important to guess the precise analytical form of $p(\pi_n)$, but is not critical: Portelli *et al* showed that two rather different analytical expressions, log-normal and $\chi$-squared, yield predictions that are hardly distinguishable-within experimental error bars-over the available range of variation of the independent variable, with a little advantage in favour of the $\chi$-squared PDF; also theoretical reasons are known since long, suggesting that the log-normal is not the ideal candidate PDF [10]. Hence, from here on, we will limit to consider Eq. (6) for $p(\pi_n)$, knowing that even departures from this form-within some limits-are not likely to give remarkable differences.

Notice that here we are speaking about PDFs for $\pi_n$ variables. They are different from the PDF for $\Delta N$, which is instead the physically relevant variable. However, it is known that log-normal PDFs are often associated with fragmentation process, hence we may expect $P(\Delta N)$, also, to be close to a log-normal curve. In the paper [12] we developed a semi-phenomenological model for density fluctuations. A model charge continuity equation was written, yielding a functional dependence between density and potential fluctuations. The latter ones had to be guessed from experiment. The result was, approximately, a lognormal form for density fluctuations which fitted well data over most of their range. It is interesting to see that the result [1] can be accommodated within the present model by choosing a log-normal form for $p(\pi_n)$, together with $\Lambda$=0.

The Fourier transform of $p(\pi_n - \mu_0)$ is, thus,

$$\tilde{p}(k) = \frac{\exp(ik\mu_0)}{\left(1 + i\frac{k\mu_0}{\nu}\right)^{\nu}} \qquad (7)$$

This expression is straightforwardly generalized to the product over the index $n$, and one gets, from Eq. (3)

$$P(\Delta N) = \int_{-\infty}^{+\infty} \frac{dk}{2\pi} \exp\left[ ik\Delta N + \sum_{\nu}\left( ik\mu_0 - \nu \log\left(1 + i\frac{k\mu_0}{\nu}\right)\right)\right]$$

$$= \frac{1}{2\pi\mu_0} \int_{-\infty}^{+\infty} d\xi \exp\left[ i\xi\left(\frac{\Delta N}{\mu_0}\right) + \sum_{\nu}\left( i\xi - \nu \log\left(1 + i\frac{\xi}{\nu}\right)\right)\right]$$

(8)

where we have written for compactness the sum over the index $\nu$, but remember that, more appropriately, it is over $n$ ($0 \leq n \leq \Lambda$), and $\nu = [\rho^n]$. Apart from a trivial normalization factor, the fitting formula (8) depends upon three free parameters: $\mu_0$, $\rho$, $\Lambda$ (or $Re$). The former two are likely to be related solely to the microscopic processes governing the turbulence. The third, playing the role of Reynolds number, should in principle depend upon the macroscopic setup as well.

## 3. Testing against RFX data

In the following, we will test our model against experimental data from RFX Reversed Field Pinch [1]. The data were taken at the very edge plasma using Langmuir probes with a sampling frequency of 1 MHz, during the flat-top phase of pulses. The total number of collected points ranges between $2\times10^4$ and $4\times10^4$. Langmuir probes are operated in RFX only with low-current low-temperature plasmas. Edge temperature is varying in the 10-30 eV range for these pulses, and is only weakly dependent upon core temperature. For the same conditions, edge density is less than $1\times10^{19}$ m$^{-3}$. More details about experimental arrangements can be found in [1].

Figure (1) is the main result of this work and features a few samples of the PDFs for density fluctuations, together with fits produced using Eq. (8). From top to bottom, we feature different probe insertions, for excursions differing up to about 1 cm. The best fitting curves-the solid ones-yield in all cases an excellent interpolation of experimental data. Error bars, of course, account for qualifying the goodness of a fit. We are not able to quantify the errors due to measurement. Limiting to those due to statistics (and assuming Poissonian statistics), we can state that they would be of the same size of the symbols as appearing in figure. The $\mu_0$ parameter has the meaning of a scaling factor and is important in deciding the slope of the PDF at high $\Delta N$'s The parameter ρ has typical value ≈ 40, yields a fraction between 3 and 4 for relative linear sizes of eddies at two successive scales. As for Λ, from the structure of the function (8) it is expected that the terms with higher ν's give smaller and smaller contributions. In Fig. (1), top plot, it is shown that the difference between retaining terms up to Λ = 3 (chain curve) or Λ = 6 (solid curve) is fairly small. On the other hand, small ν's gauge the importance of the departure of self-similarity, since this is expected to be more relevant for small scales, close to the dissipative ones: let us imagine, in the sum (8), to remove the lowest terms, leaving only higher ν's. This corresponds to imposing self-similarity at the smallest scales. The result is that the PDF *P(ΔN)* approaches a Gaussian. Hence, this stresses again that departures from self-similarity are essential to recover experimental data. The same result was found on RFX through a wavelet analysis of experimental data. Wavelets filtering allows to discern contributions from differing time scales (or space scales, if Taylor's frozen turbulence hypothesis holds). At the largest scales, all PDF's were found to converge towards Gaussians, while departures from Gaussians became more and more relevant towards smaller scales [4].

The issue of departure from self-similarity has been pointed out by several papers to be a key feature of plasma edge turbulence [4]. This fact was often used negatively, i.e., to rule out some models as unsuitable candidates for the description of turbulent transport. In this work we were able to use it a key element in a constructive fashion, as an ingredient within a micoscopical plasma modelling. The physical process we have built is basically a *direct* transfer of *particle density,* from larger to smaller scales. The issue of the existence of direct or inverse cascades is still an open-ended question in turbulence, both for neutral fluids as well as for plasmas. It is known, by example, that two-dimensional Magneto Hydrodynamics (MHD) predicts an *inverse* cascade process for *energy*. The Reynold stress is the term in fluid or MHD equations that may interchange energy between different scales in plasmas. Indeed, theoretical speculations do exist since long supporting the idea that both kinds of cascades may exist in plasmas-possibly depending on scales [1]. Experimentally, only few partial results still do exist, aimed at investigating the effects of this term. The evidence coming from them, at present, is that such energy transfers do actually occur [1] and that may be functions not only of spatial scale but also of the position into the plasma [1]. Of course, since no straightforward correspondence may be envisaged between energy and density behaviour, this can give just an insight of what may be expected for density. Hence, the whole question whether in this work we have been describing a mechanism truly at work in plasmas is largely unanswered, although the good agreement obtained here may seen as an hint in favour of the existence of direct cascades.

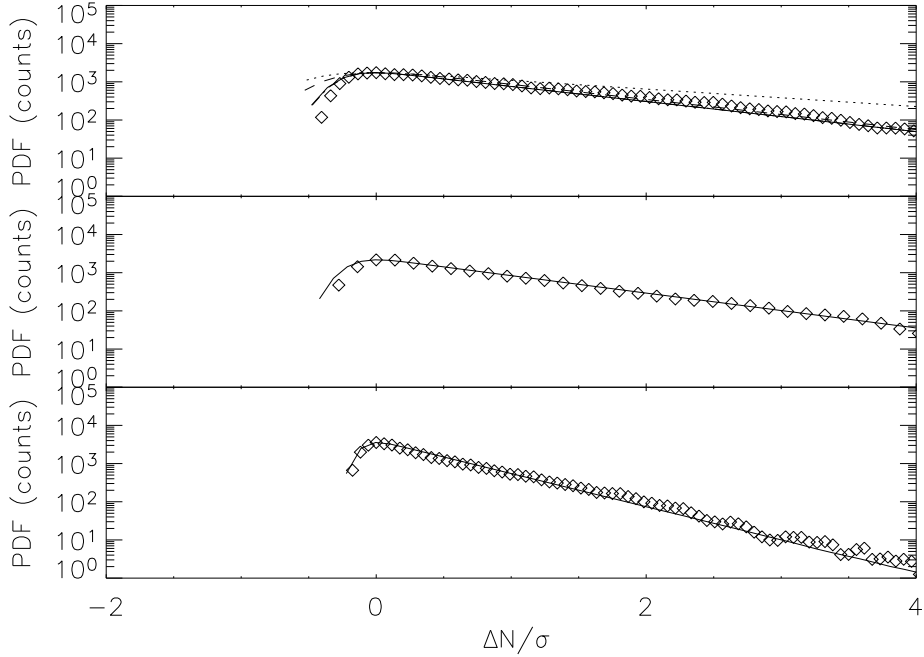

Figure 1. Symbols, PDF of experimental density fluctuations. From top to bottom, deeper to shallower probe insertion. The density is normalized to the experimental mean square deviation, and shifted by an offset so that the maximum of the PDF is for $\Delta N = 0$. Statistical error bars are about the same size of the symbols. Curves, Eq. (8) for various values of parameters $\mu_0$, $\rho$, $\Lambda$. Top plot: solid curve ($\mu_0 =1.1$, $\rho = 45$, $\Lambda= 6$); dashed curve ($\mu_0 =1.1$, $\rho = 17$, $\Lambda= 6$); dotted curve ($\mu_0 =1.8$, $\rho = 45$, $\Lambda= 6$); chain curve ($\mu_0 =1.1$, $\rho = 45$, $\Lambda= 3$). Chain curve is almost perfectly overlapping the solid curve. Middle plot: ($\mu_0 =0.95$, $\rho = 45$, $\Lambda= 6$). Because of the smaller number of counts, in order to keep statistics constant, this plot has been generated using lesser bins. Bottom plot: ($\mu_0 =0.5$, $\rho = 35$, $\Lambda= 6$).

## 4. Concluding remarks

Summarizing, we think we have been able to give a fairly good account of phenomenology within a single interpretive framework. Up to now, our approach has led practically to a fitting formula containing some arbitrary parameters that are just fixed by matching PDF data. It would be reassuring being able to relate the numerical values of these parameters with corresponding quantities actually found in plasmas. Let us try to do this step further: the most straightforward quantities to be dealt with are $L$ and $\eta$, the characteristic length scales involved. An intuitive meaning for $L$ is the typical scale at which coherent structures are observed, that is order of centimeters. A bit more difficult is to attach a meaning to $\eta$; however, a lower bound for it is naively found: $\eta$ cannot be smaller than Debye length $\lambda_D$, since at this length the fluidlike description of plasma must break down. Hence, we may assume $\eta \geq \lambda_D$. From previous paragraphs, and using typical values for $\rho$, $\Lambda$ as arising from figure 1, we get $\eta/L = \rho^{-\Lambda/3} \leq 10^{-3}$, or $\eta \leq 10^{-5}$ m. This is comforting since-in RFX- $\lambda_D \approx O(10^{-6 \div -5})$ m.

To finish with, we address another important issue, correlated with the present work: a considerable interest has been raised in the past (and still is) in searching for unifying features from disparate turbulent systems, independently from specific models. Undisclosing universal aspects, common to all or to a class of turbulent systems, may shed light on the underlying physics, when lacking better information. Again, we limit here to statistical tools dealing with PDFs. In recent years, some interesting works appeared concerning universal features of PDFs in several *strongly correlated* systems [1]. The suggestion, there, was that PDFs of fluctuations follow just one universal curve (BHP curve), a generalized form of the Gumbel's distribution $G_a(x)$. Gumbel's curve is well known in statistics, giving the probability of picking the $a$-th largest value from an ensemble of *uncorrelated* variables. Connections between these systems and Extreme Value Distributions, hence, arise intuitively. BHP curve is peculiar in that $a$ is universal and noninteger: $a = \pi/2$. It was suggested this value to represent an effective number of degrees of freedom for a system of correlated variables. Indeed, BHP curves apparently strongly resemble the skewed distributions we have found in Fig. (1). Hence, the question whether BHP approach could be extended to our plasmas appears fairly interesting[1].

As far as we understand, however, the matter is not still entirely settled: there are claims that BHP functions could not be truly universal [20]; also, Watkins *et al* [21] pointed out that long-range correlations could not be the only ingredient leading to BHP curve, but also the finite-size of the system is. Finally, Rypdal and Ratynskaia [22] carried on recently an analysis of fluctuations in a magnetized (but not fusionist) plasma using, among other tools, the BHP approach. Although their results were encouraging, they commented that, unless one has a very good statistics spanning long intervals, there are several possible distributions fitting the data within approximately the same accuracy. Hence, no definite claim may be made of the superiority of one distribution over the others (We made a similar comment in the previous paragraphs after Eq. 6). Keeping in mind this caveat, we performed a fit of data in Fig. (1) using generalized Gumbel distributions, but leaving $a$ as a free parameter. Our results do not appear supporting BHP distributions: the accuracy imposed by our data was enough to definitely rule out the possibility $a = \pi/2$, while we found rather that a reasonable fit of data was obtained only for $a \approx 0.1$. Chapman *et al* [23] already showed that $a$ must be a system-dependent parameter, and hence may depart slightly from BHP value (see, about this, also the paper by Noullez and Pinton in [1]), but we are not aware of any satisfactory interpretation of this parameter that may accommodate values lesser than unity.

---

[1] While this paper was finishing the reviewing stage, the paper [19] appeared, where Van Milligen *et al* tackled the same kind of analysis in tokamak plasmas.